\documentclass[
aps,prl,twocolumn,footinbib,
raggedbottom,superscriptaddress,altaffilsymbol,amsmath,amssymb,amsfonts]
{revtex4-1}

\usepackage{graphicx}
\usepackage{bm}
\usepackage{url}

\begin{document}

\title{Radiative deflection by spin effect\\
in the quantum radiation-reaction regime}

\author{X. S. Geng}
    \affiliation{State Key Laboratory of High Field Laser Physics, Shanghai Institute of Optics and Fine Mechanics, Chinese Academy of Sciences, Shanghai 201800, China}
    \affiliation{University of Chinese Academy of Sciences, Beijing 100049, China}
    
\author{L. L. Ji}
    \email{jill@siom.ac.cn}
    \affiliation{State Key Laboratory of High Field Laser Physics, Shanghai Institute of Optics and Fine Mechanics, Chinese Academy of Sciences, Shanghai 201800, China}

\author{B. F. Shen}
    \email{bfshen@mail.shcnc.ac.cn}
    \affiliation{Shanghai Normal University, Shanghai 200234, China}
    \affiliation{State Key Laboratory of High Field Laser Physics, Shanghai Institute of Optics and Fine Mechanics, Chinese Academy of Sciences, Shanghai 201800, China}
    
\author{B. Feng}
    \affiliation{State Key Laboratory of High Field Laser Physics, Shanghai Institute of Optics and Fine Mechanics, Chinese Academy of Sciences, Shanghai 201800, China}
    \affiliation{University of Chinese Academy of Sciences, Beijing 100049, China}
    
\author{Z. Guo}
    \affiliation{State Key Laboratory of High Field Laser Physics, Shanghai Institute of Optics and Fine Mechanics, Chinese Academy of Sciences, Shanghai 201800, China}
    \affiliation{University of Chinese Academy of Sciences, Beijing 100049, China}
    
\author{Q.Q. Han}    
    \affiliation{State Key Laboratory of High Field Laser Physics, Shanghai Institute of Optics and Fine Mechanics, Chinese Academy of Sciences, Shanghai 201800, China}
    \affiliation{University of Chinese Academy of Sciences, Beijing 100049, China}
\author{C.Y. Qin}
    \affiliation{State Key Laboratory of High Field Laser Physics, Shanghai Institute of Optics and Fine Mechanics, Chinese Academy of Sciences, Shanghai 201800, China}
    \affiliation{University of Chinese Academy of Sciences, Beijing 100049, China}
\author{N.W. Wang}
    \affiliation{State Key Laboratory of High Field Laser Physics, Shanghai Institute of Optics and Fine Mechanics, Chinese Academy of Sciences, Shanghai 201800, China}
    \affiliation{University of Chinese Academy of Sciences, Beijing 100049, China}
\author{W.Q. Wang}
    \affiliation{State Key Laboratory of High Field Laser Physics, Shanghai Institute of Optics and Fine Mechanics, Chinese Academy of Sciences, Shanghai 201800, China}
    \affiliation{University of Chinese Academy of Sciences, Beijing 100049, China}
\author{Y.T. Wu}
    \affiliation{State Key Laboratory of High Field Laser Physics, Shanghai Institute of Optics and Fine Mechanics, Chinese Academy of Sciences, Shanghai 201800, China}
    \affiliation{University of Chinese Academy of Sciences, Beijing 100049, China}
\author{X. Yan}
    \affiliation{State Key Laboratory of High Field Laser Physics, Shanghai Institute of Optics and Fine Mechanics, Chinese Academy of Sciences, Shanghai 201800, China}
    \affiliation{University of Chinese Academy of Sciences, Beijing 100049, China}
\author{Q. Yu}
    \affiliation{State Key Laboratory of High Field Laser Physics, Shanghai Institute of Optics and Fine Mechanics, Chinese Academy of Sciences, Shanghai 201800, China}
    
\author{L.G. Zhang}
    \affiliation{State Key Laboratory of High Field Laser Physics, Shanghai Institute of Optics and Fine Mechanics, Chinese Academy of Sciences, Shanghai 201800, China}
    \affiliation{University of Chinese Academy of Sciences, Beijing 100049, China}
    
\author{Z.Z. Xu}
    \affiliation{State Key Laboratory of High Field Laser Physics, Shanghai Institute of Optics and Fine Mechanics, Chinese Academy of Sciences, Shanghai 201800, China}
    \affiliation{University of Chinese Academy of Sciences, Beijing 100049, China}

\date{\today}

\begin{abstract}
The colliding between ultra-relativistic electrons and an ultra-intense laser pulse is a powerful approach to testify the physics in strong-field QED regime. By considering spin-dependent radiation-reaction during laser-electron collision we find anti-symmetric deflection of electrons with different spin states. 
We revealed that such deflection is induced by the non-zero work done by radiation-reaction force along field polarization-direction in a half-period of phase, which is larger for spin-anti-paralleled electrons and smaller for spin-paralleled electrons. 
The spin-projection on the magnetic field of an electron gets inversed in adjacent half-periods due to oscillating magnetic field and therefore the deflection due to spin-dependent radiation is accumulated rather than vanishing. 
The new mechanism provides an extra dimension to observe quantum radiation-reaction effect in the strong-field QED regime by measuring the anti-symmetric distribution.
\end{abstract}


\maketitle
The rising laser intensity enables us of investigating the dynamics of electrons in intense field,  e.g. in fast ignition fusion, laser-driven particle acceleration and bright X/gamma-ray sources. 
When laser intensity approaches $10^{23}$ $\text{W/cm}^2$ electron dynamics is strongly coupled with photon radiation and consequent recoil effect, which is usually referred to as the radiation-reaction (RR) effect.
Such phenomenon can be further enhanced by colliding the laser with high energy electrons where RR becomes dominant as the field strength in the rest frame of the electron is boosted by a factor of $\sim\gamma$ (the gamma factor of a relativistic electron) \cite{LL}.

In the QED perspective, when the spin state of an electron is considered it has been found out that the spin-anti-paralleled electron tends to radiate more energy than the spin-paralleled electron \cite{Ternov1995,seipt-radiative_pol-pra2018}.
In the radiation-reaction regime, spin-flip along with ``quantum-jump'' process were proposed to polarize an unpolarized electron beam in the collision with an elliptically polarized laser pulse, which is not possible for linearly polarized laser \cite{liyanfei-arxiv2018}. 
However, in cases where spin-flip rate is negligible, when passing through the oscillating laser field the spin projection on the magnetic field axis in the rest frame of electron oscillates with the field and the net contribution from spin is averaged to zero.
Therefore it seems that no sign of spin-dependent dynamics would emerge in the laser-electron collision, considering that the Stern-Gerlach effect is too weak to play a role \cite{wenmeng-pra2017} (will be discussed later).

In this article, we will show that when radiation-reaction is coupled with spin dynamics, spin-dependent radiation causes accumulative deflection of polarized electrons and causes inhomogeneous polarity after colliding laser with an unpolarized electron bunch. The distinctive spin-dependent signature in electron dynamics can be testified in the next-generation 10-100 PW laser facilities \cite{SEL, ELIweb, XCELSweb, Apollonweb, Vulcanweb, SULF} to account for quantum RR in the strong-field QED regime,
as the change of polarity provides a new degree of freedom in addition to gamma photon emission \cite{wistisen-NC2018}, electron energy reduction \cite{cole-PRX2018,poder-PRX2018} and electron motion \cite{Quenching,Geng-arxiv2018}.
To our knowledge the mechanism of the spin-dependent radiative deflection has not been revealed previously.

    We mimic the collision process via a QED Monte-Carlo (MC) modelling as well as the classical equation of motion.
    The radiation power for spin-paralleled/-anti-paralleled electrons was given in Ref. \cite{Ternov1995}. The theory for arbitrarily polarized electrons is further developed in Ref. \cite{seipt-radiative_pol-pra2018}. For electron polarization vector of $\bm{s}$ in its rest frame, the radiation probability rate is \cite{seipt-radiative_pol-pra2018}
    \begin{equation}\label{eq:section probability}
        \frac{dP}{d\delta d\psi}=-\frac{\alpha}{b}\left[
        Ai_1(z)+g\frac{2Ai'(z)}{z}+s_\zeta \delta\frac{Ai(z)}{\sqrt{z}}
        \right]
    \end{equation}
    where $\alpha$ is the fine structure constant, 
    $b=\hbar k\cdot p/m^2c^2$, $Ai(z)$ the Airy function, 
    $\delta=\hbar k\cdot \hbar k'/\hbar k\cdot p$, 
    $\psi$ the laser phase,
    $\hbar k$, $\hbar k'$ and $p$ the four-momentum of laser, photon and electron,  
    $z=[\frac{\delta}{(1-\delta)\chi_e}]^{2/3}$, 
    $g=1+\delta^2/[2(1-\delta)]$, 
    $\chi_e=e\hbar|F\cdot p|/m^3c^4$, 
    $F$ the electromagnetic tensor, 
    $e$ the electron charge, 
    $\hbar$ the reduced Planck constant, 
    $m$ the electron mass at rest, 
    $c$ the speed of light 
    and $s_\zeta=\bm{s\cdot \hat{B}}_\text{rest}$, respectively. 
    Here $\bm{\hat{B}}_\text{rest}$ is the direction of the magnetic field in the electron's rest frame. 
    The dependence on spin-orientation vanishes in the collision with a symmetric laser field. 
    For spin-flip events, the radiation probability for $dP_{\text{flip}}/d\delta d\psi$ and $dP_{\text{nonflip}}/d\delta d\psi$ are given in Ref. \cite{seipt-radiative_pol-pra2018} 
    where the electron undergoes $\bm{s}\rightarrow-\bm{s}$ in a spin-flip process.

    We numerically solve the equation of motion $d\bm{p}/dt=-e(\bm{E}+\bm{p}\times\bm{B}/\gamma m)$ between photon emission events,  
    where the latter is modelled via QED-MC algorithm \cite{PICQED}. 
    Random numbers are generated to sample the radiation spectrum at each time step to determine an emission event and the photon energy. 
    The first random number defines the probability of radiation; 
    the second random number determines whether the radiation happens; 
    then the spin-flip event is determined by the weights of flip and non-flip probabilities.

    For classical description, we treat RR as the average of quantum photon emission.
    The classical RR is described by the Landau-Lifshitz equation \cite{LL} which is quantum-corrected by multiplying the classical radiation power with the spin-dependent Gaunt factor \cite{sorbo-pol-ppcf2018}
    \begin{multline}\label{eq:gaunt}
        g^s=[
        1+2.54(\chi_e^2-1.28s\chi_e)+(4.34+2.58s)(1+\chi_e)\\
        \times ln(1+(1.98+0.11s)\chi_e)
        ]^{-2/3}
    \end{multline}
    where the spin is in parallel or anti-parallel to the chosen axis $\bm{B}_\text{rest}$. In ultra-relativistic limit, 
    $\bm{F}_\text{RR}^s\sim g^s a^2 \gamma^2$. 
    The contribution from spin-flip process is ignored here due to low flip-rate which will be discussed later.
    
    When QED radiation is averaged to classical radiation, the stochastic quantum effect is then omitted. 
    In QED, electrons do not necessarily radiate the same amount of energy as they do classically due to the stochastic photon emission  \cite{dipiazza-stochastic-prl2013,vranic-qrr-njp2016,Quenching,lijianxing-angle_resolved_pho-scirep2017,ridgers-signature_qrr-2017}. In our case the energy loss for spin $\pm$ electrons can be well interpreted by its averaged representation, which successfully accounts for the spin-dependent radiative deflection.
    
    As for the Stern-Gerlach force, one can estimate it through $\bm{F}_{SG}=q/m\bm{\nabla}(\bm{s}\cdot\bm{B}_\text{eff})$
    ,where $\bm{B}_\text{eff}=\bm{B}/\gamma-\bm{p}\times\bm{E}/\gamma mc^2/(\gamma+1)$ \cite{wenmeng-pra2017}. 
    The Stern-Gerlach force can also deflect electrons of different spin by the divergence of magnetic field in a laser-electron collision. 
    However, the difference between the deflection angle of spin $\pm$ is at the order of ${10}^{-7}$ rad when $\gamma_0=100,a_0=200$ \cite{wenmeng-pra2017}, which is negligible comparing to the radiative deflection by spin effect.
    Therefore it is not considered throughout this article.
    
    In our modelling, the laser propagates along the $z$-axis, where $\hat{\bm{x}}$ is the E-field direction and $\hat{\bm{y}}$ is the B-field direction. Electron spin is defined in this coordinate with a unit vector, e.g. $(0,\pm 1,0)$ indicates the spin is parallel/anti-parallel to the $y$-axis.

Head-on collision between polarized electrons and a laser is firstly investigated using the QED-MC method.
We start with the simplest situation where the laser is approximated by 
\begin{equation}\label{eq:pulse}
    \bm{E}=\hat{\bm{x}}E_0 exp(-x^2/w_0^2)sin(\psi/2N)^2cos(\psi)
\end{equation}
and $\bm{B}=\hat{\bm{y}}E_x/c$ with the wavelength of 800 nm.
Here $E_0$ is the electric field amplitude, $w_0$ the beam waist at $e^{-1}$, $\psi=\omega t-kz$ the phase, $N$ the pulse length measured by wavelength, respectively. 
Eq. (\ref{eq:pulse}) is a good approximation here since the electrons pass only the near-axis part of the laser beam.
Electrons are initially polarized along $\pm y$ so that the spin-vector does not precess.
We choose $y$-axis as the reference axis for electron spin. 
In the head-on collision setup and plane-wave approximation, the $\hat{\bm{B}}_\text{rest}$ is consistent with the magnetic field the laboratory frame because $\bm{B}_\text{rest}
=\gamma\bm{B}-\frac{\bm{p}}{mc^2}\times\bm{E}-\frac{\bm{(p\cdot B)p}}{(\gamma+1)mc^2}
=\hat{\bm{y}}(\gamma-\sqrt{\gamma^2-1}\beta_z)B_\text{lab}(\psi,\bm{r})$
where $\bm{\beta}=\bm{v}/c$. Therefore $\bm{s}\cdot\hat{\bm{B}}_\text{rest}$ oscillates with laser phase.

\begin{figure}
    \includegraphics[width=0.48\textwidth]{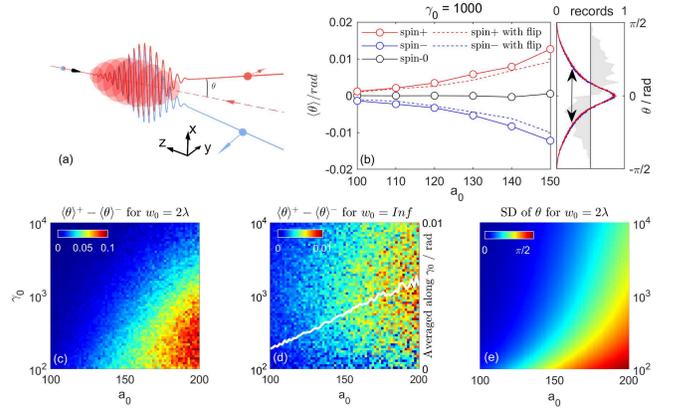}
    \caption{
    (a) Collision between a polarized electron and a strong laser pulse. The electron propagates along $-z$ direction; the laser propagates along $z$ direction with polarization along $x$. Electrons of different initial spin polarization, i.e. parallel/anti-parallel, get deflected to opposite direction due to spin-dependent radiation-reaction. 
    (b) The deflection angle for different field strength with $\gamma_0=1000$. Curves on the right are the angular distribution of $\pm$ polarized electrons after collision for $a_0=150$ and the gray area is the difference between spin $\pm$. Electron records are normalized in all cases.
    (c-d) The difference between $\left\langle\theta\right\rangle^\pm$ for $w_0=2\lambda\text{and Inf}$ in the $a_0-\gamma_0$ parametric space. The white line in (d) is the average along $\gamma_0$.
    (e) The standard deviation of deflection angles which reflects the spread of the scattered electron shown by the double-arrow in (b).
    }
    \label{fig1}
\end{figure}

We find that electrons of opposite spin-orientation ($\pm$) tend to be scattered to opposite directions (Fig. \ref{fig1}(a)) via QED-MC. 
For reasons to be shown later, one will find more electrons of parallel/anti-parallel polarization in the upper ($\theta>0$) / lower ($\theta<0$) region, resulting in slight asymmetry of electron distribution along $\theta$ in Fig. \ref{fig1}(b). 
We qualify such asymmetric distribution by defining the averaged deflection angle $\langle\theta\rangle$. 
The phenomenon is investigated in a large parameter regime as shown in Fig. \ref{fig1}(b-e).
We will focus on the $\chi_e < 1$ region, e.g. Fig. \ref{fig1}(b) where pair-production \cite{BreitWheelerPP} is suppressed \cite{Ritus}. 
The averaged deflection angle is calculated by repeating the collision for ${10}^6$ times in Fig. \ref{fig1}(b) and ${10}^5$ times in Fig. \ref{fig1}(c-e) for each set of parameters. 
For fixed electron energy, larger field strength can separate electrons of different initial spin-orientation to larger angles, as shown in Fig. \ref{fig1}(b), while the deflection angle of spin-free electrons stays near zero as $\gamma_0\gg a_0$. 
The deflection angle is slightly reduced when spin-flip is turned on due to the spin-flipped electrons tending to be scattered to the opposite direction. 
Since every electron averagely flips only 0.18 time at $a_0=150, \gamma_0=1000$ according to our modelling, which is minor as shown in Fig. \ref{fig1}(b). We therefore will not include this effect in our following discussion. 
The deflection angle can be further enhanced for larger $a_0$ and smaller $\gamma_0$ as shown in Fig. \ref{fig1}(c).
In fact, it is because the deflection is significantly enhanced by ponderomotive scattering, which is more effective for the lower energy electrons and higher laser fields. 
It can be further verified by the coincidence between Fig. \ref{fig1}(c) and \ref{fig1}(e) where the standard deviation of $\theta$ reflects the spread of scattered electrons as well as the strength of ponderomotive scattering, 
which indicates the role of ponderomotive scattering in spin-dependent deflection.
It should be noted that when ponderomotive scattering is turned off (with infnite beam waist), as shown inf Fig. \ref{fig1}(d), the deflection angle scales linearly with $a_0$ and weakly depends on $\gamma_0$.

\begin{figure}
    \includegraphics[width=0.48\textwidth]{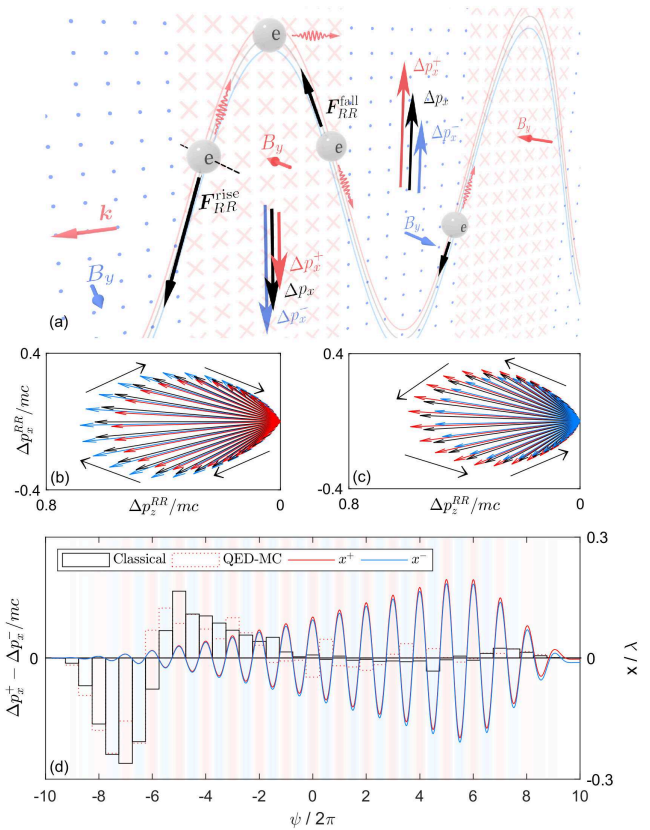}
    \caption{
    An electron collides with a flat-top laser pulse where the plateau region is $-5<\psi/2\pi<5$ for spin $+/-$ (red/blue) and spin-free (black).
    (a) Electron trajectories (curves) and momentum change due to RR in $x$-direction in half-period ($\Delta p_x$ and arrows). The lengths of arrows indicate the length of the vector. 
    (b-c) The variation of vector $\Delta p^{RR}$ along the collision time steps for adjacent half-periods (``$\times$'' for (b), ``$\cdot$'' for (c)). Integrate the vectors along $x$ and one will get $\Delta p_x^\pm$ shown in (a). 
    (d) The electron trajectories of initially $+/-$ polarized electron (red/blue lines) and the difference between $\Delta p_x^\pm$ accumulated in half-periods for classical radiation (black bars) and QED-MC radiation (red-dotted bars).
    }
    \label{fig2}
\end{figure}

One can find in Eq. (\ref{eq:section probability}) or in Eq. (\ref{eq:gaunt}) that electrons radiate more energies when anti-paralleled to $\bm{B}_\text{rest}$. 
As a result, an electron experiences larger (anti-paralleled) or smaller (paralleled) $\bm{F}_\text{RR}$ compared to a spin-free electron. 
To reveal the mechanism, we now focus on the ponderomotive oscillation of an electron during its collision with a plane-wave as shown in Fig. \ref{fig2}(a). 
When RR force is included, since $\bm{F}_\text{RR}\sim\gamma^2$ the electron loses its energy during oscillation such that the work done by RR force towards $x$-direction cannot be fully compensated by the one along the $-x$-direction, 
i.e. $\left|\int_\text{rise}{d\psi\bm{F}_\text{RR}\cdot\hat{\bm{x}}\ }\right| > \left|\int_\text{fall}{d\psi\bm{F}_\text{RR}\cdot\hat{\bm{x}}}\right|$. 
Thus the net work is non-zero in the half-period. 
As seen in Fig. \ref{fig2}(a), the concavity of the trajectory curve is locked with the phase as well as $\hat{\bm{B}}_\text{rest}$. 
Therefore, the damping force induces net negative/positive $\Delta p_x^{RR}$ \footnote{$\Delta p_x$ is used for convenience.} (the momentum shift due to RR force) for the concave-down/-up curve shown by the down/up black arrows in Fig. \ref{fig2}(a). 
In this situation electron with spin (anti-)parallel to $\bm{B}_\text{rest}$ ($\bm{B}_\text{lab}$ equivalently) experiences a (larger) smaller $\Delta p_x$ compared to the spin-free case, as shown by the length of the arrows (red for $+$, blue for $-$, black for spin-free). 
For spin-parallel electron in the concave-down curve, the spin state becomes anti-parallel in the next concave-up curve. 
As a result, by integrating $\Delta\bm{p}^\pm$ along $x$-direction, shown by the momentum shift $\Delta\bm{p}^{RR}$ at each time step in Fig. \ref{fig2} (b-c), we find that the electron always experiences a net up-shift (sum of red arrows) or a net down-shift (sum of blue arrows) relative to the spin-free electron (black arrows), 
i.e. $\Delta p_{x,\text{total}}^+>\Delta p_{x,\text{total}}>\Delta p_{x,\text{total}}^-$, during the half-periods, which can also be concluded from the asymmetry along $x$-axis in Fig. \ref{fig2} (b-c). 
Therefore the momentum shift of adjacent half-periods does not cancel each other and the deflection accumulates.
The momentum change will later be confirmed by our numerical results. 
From the above analysis, one can see that the deflection direction relative to spin-free electron can be defined by $\bm{s}\times\bm{k}$ in the plane-wave approximation.

The ponderomotive scattering and pulse shape contribute to the deflection angle. 
To reveal the net effect by spin-dependent radiation, we perform the collision for a flat-top pulse without beam waist. 
We focus on the plateau ($-5<\psi/2\pi<5$) to exclude the effect of the rising/falling edge. 
Here the classical equation of motion with spin-dependent Gaunt factor is employed so that the time-resolved dynamics are not influenced by stochastic effects from QED-MC. 

The mechanism shown by Fig. \ref{fig2}(a) is directly proved in Fig. \ref{fig2}(d) by colliding electron of $\gamma_0=1000$ with flat-top pulse of $a_0=150$ and infinite beam size, 
where the difference between $\Delta p_x^\pm$ in each half-period are presented (black and red-dotted bars). 
One can see $\Delta p_x^+$ becomes greater than $\Delta p_x^-$ after the plateau begins ($\psi/2\pi>-5$) and the electron is deflected due to such momentum change. 
The deflection direction is along the $\bm{s}\times\bm{k}$ axis, which is exactly as predicted with our model in Fig.\ref{fig2}(a) in the plane-wave approximation. 
We note that in the rising edge of the pulse ($-10<\psi/2\pi<-5$), the momentum change from RR is inversed, $\Delta p_x^+<\Delta p_x^-$, as shown by the negative black bars.
In this case, the increase of the field strength in the rising edge dominates over the correction from spin-dependent emission in terms of the RR force,
therefore $\bm{F}_\text{RR}^\text{fall}$ overrides $\bm{F}_\text{RR}^\text{rise}$.
The net work in $x$-direction is opposite to that in the plateau region
$\left|\int_\text{fall}{d\psi\bm{F}_\text{RR}\cdot\hat{\bm{x}}\ }\right| > \left|\int_\text{rise}{d\psi\bm{F}_\text{RR}\cdot\hat{\bm{x}}\ }\right|$. 
Accordingly, in the falling edge we have $\Delta p_x^+>\Delta p_x^-$ again. 
The counter-play between the field strength variation and the damping of $\gamma$ can be clearly seen from $\bm{F}_\text{RR}\sim a^2\gamma^2$. 

\begin{figure}
    \includegraphics[width=0.48\textwidth]{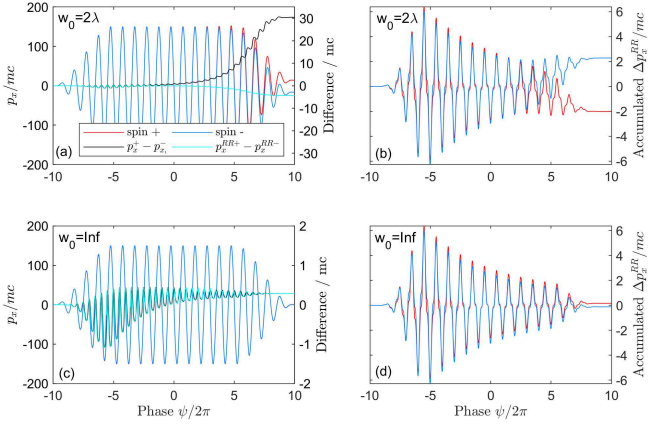}
    \caption{
    (a, c) The momentum change of $\pm$ polarized electrons (red/blue line) during collision with a flat-top laser with $w_0=2\lambda$ and infinite beam waist. The difference between $p_x^\pm$ is shown by the black line. 
    (b, d) Accumulated momentum change from RR of $\pm$ polarized electrons, i.e. $p_x^{RR\pm}$ (red/blue line). The difference between $p_x^{RR\pm}$ is shown by the cyan line in (a) \& (c).
    }
    \label{fig3}
\end{figure}

For further analysis of the ponderomotive effect, the electron momentum and momentum change from radiation are presented in Fig. \ref{fig3} by colliding the electron with a flat-top laser of beam waist $w_0=2\lambda$ and infinite beam waist. 
In Fig. \ref{fig3}(a) and (c), one can see that $p_x$ oscillates with constant amplitude during the plateau while the difference between $+$(red) and $-$(blue) gradually builds up (black). 
Such difference grows much quicker with the help of ponderomotive force ($\sim-\nabla E^2$) by comparing Fig. \ref{fig3}(a) to (c). Fig. \ref{fig3}(b) and (d) shows the accumulated momentum change due to $\bm{F}_\text{RR}$ along the $x$-direction, i.e. $p_x^{RR\pm}$; 
the differences between $+$(red) and $-$(blue) are presented in Fig. \ref{fig3}(a) and (c) (cyan).
The coincidence between the black and cyan lines in Fig. \ref{fig3}(c) indicates that the momentum loss from radiation is the only source of electron deflection in plane-wave.
One the other hand, the difference between the black and cyan lines in Fig. \ref{fig3}(a) suggests the deflection is further amplified by ponderomotive scattering. 
The deflection angle becomes large enough due to ponderomotive enhancement such that an electron radiates more energy upwards/downwards if it is going up/down, as seen from the flipped momentum loss in Fig. \ref{fig3}(b).

    \begin{figure}
        \includegraphics[width=0.48\textwidth]{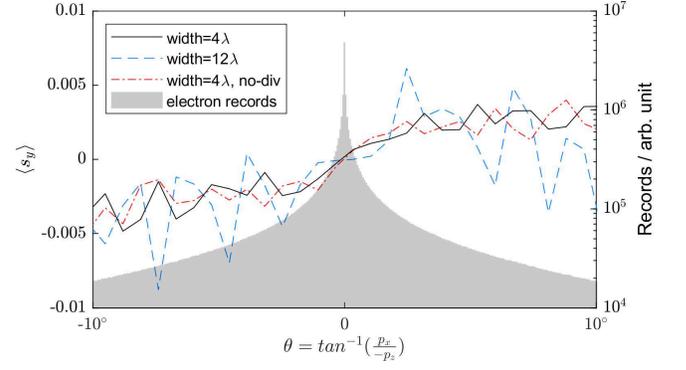}
        \caption{
        Angular distribution of the unpolarized electron bunch of $\gamma_0=1000$ after collision with a laser pulse of $a_0=150$ (gray area) 
        and the angular distribution of $\langle s_y\rangle$ for small transverse size of $4\lambda$ with 1\% energy spread and 10 mrad angular divergence (black-solid), 
        large transverse size of $12\lambda$ (blue-dashed) 
        and transverse size of $4\lambda$ without energy and angular divergence (red-dot-dashed). 
        }
        \label{fig4}
    \end{figure}
    
    Now that the electron with opposite spin-orientation tend to be scattered to opposite directions in a collision with laser pulse, the averaged spin polarization therefore becomes inhomogeneous as a function of the scattered angle. 
    We consider an unpolarized electron bunch of $\gamma_0=1000$ with transverse size of $4\lambda$ and $12\lambda$ at $e^{-1}$ with energy spread of $\sim1\%$ and angular divergence of 10 mrad colliding with a laser pulse of $a_0=150$, 
    then the polarity of electrons is measured along $y$-direction for different scattering angle. 
    The radiation is treated in QED-MC manner without spin-flip due to the low flip-rate ; we use a focused laser pulse \cite{SalaminTightlyfocused} with FWHM of 27fs and beam waist of $2\lambda$ at $e^{-1}$. 
    In a focused pulse the spin precession needs to be considered. Spin-vector precession is calculated by solving the Thomas-Bargmann-Michel-Telegdi equation \cite{Jackson} 
    $
    d\bm{s}/dt=\frac{e}{mc}\bm{s}\times
    [
    (\frac{g_e}{2}-1+\frac{1}{\gamma})\bm{B}
    -(\frac{g_e}{2}-1)\frac{\gamma}{\gamma+1}(\bm{\beta\cdot B})\bm{\beta}
    -(\frac{g_e}{2}-\frac{\gamma}{\gamma+1})\bm{\beta\times E}
    ]
    $
    , where $a_e=\frac{g_e-2}{2}\approx1.16\times{10}^{-3}$ is the anomalous magnetic moment of electron \cite{schwinger-anomalous}.
    The inhomogeneous polarization along the $y$-axis is presented in Fig. \ref{fig4}. 
    The bunch get polarized along y because the electrons with positive $s_y$ tend to be scattered upwards while those with negative $s_y$ downwards. 
    Experimental measurement of this antisymmetric phenomenon could provide an extra dimension to observe quantum RR effect in addition to QED gamma photon emission \cite{wistisen-NC2018}, electron energy reduction \cite{cole-PRX2018,poder-PRX2018} and quantum electron motion \cite{Quenching,Geng-arxiv2018}. 
    Such phenomenon is not significantly disturbed by the energy spread and angular divergence of the electron bunch at large angles ($>$ angular divergence). 
    One can always observe the inhomogeneous polarization by comparing the red-dot-dashed to the black-solid lines in Fig. \ref{fig4}.
    However, large transverse size of the electron bunch could wipe out the signal of $\left\langle s_y\right\rangle$ at small angles ($\sim$ angular divergence) shown by the blue-dashed line in Fig. \ref{fig4} because the electrons not interacting with the laser lower the $\left\langle s_y\right\rangle$ value at small angles.
    
In conclusion, we investigate the collision of initially polarized electron and laser pulse and find the deflection due to spin-dependent radiation-reaction. 
We model such deflection with non-zero momentum loss during half-period where the momentum losses divert between parallel and anti-parallel spin-state. 
Such momentum shift is non-vanishing during the collision because if an electron radiating more upwards during some half-period, it will radiate less downwards during next half-period, which induces electron to be deflected downwards, and vice versa. 
The spin-dependent deflection causes inhomogeneous polarity after collision between an unpolarized electron bunch and an intense laser pulse, which could provide an alternative to observing the quantum radiation-reaction in the strong-field QED regime.

\begin{acknowledgments}
This work is supported by the National Science Foundation of China (Nos. 11374317), the Strategic Priority Research Program of Chinese Academy of Sciences (Grant No. XDB16000000) and the Ministry of Science,  Technology of the People's Republic of China (Grant Nos. 2016YFA0401102 and 2018YFA0404803) and the Recruitment Program for Young Professionals.
\end{acknowledgments}


\begin{thebibliography}{28}%
    \makeatletter
    \providecommand \@ifxundefined [1]{%
     \@ifx{#1\undefined}
    }%
    \providecommand \@ifnum [1]{%
     \ifnum #1\expandafter \@firstoftwo
     \else \expandafter \@secondoftwo
     \fi
    }%
    \providecommand \@ifx [1]{%
     \ifx #1\expandafter \@firstoftwo
     \else \expandafter \@secondoftwo
     \fi
    }%
    \providecommand \natexlab [1]{#1}%
    \providecommand \enquote  [1]{``#1''}%
    \providecommand \bibnamefont  [1]{#1}%
    \providecommand \bibfnamefont [1]{#1}%
    \providecommand \citenamefont [1]{#1}%
    \providecommand \href@noop [0]{\@secondoftwo}%
    \providecommand \href [0]{\begingroup \@sanitize@url \@href}%
    \providecommand \@href[1]{\@@startlink{#1}\@@href}%
    \providecommand \@@href[1]{\endgroup#1\@@endlink}%
    \providecommand \@sanitize@url [0]{\catcode `\\12\catcode `\$12\catcode
      `\&12\catcode `\#12\catcode `\^12\catcode `\_12\catcode `\%12\relax}%
    \providecommand \@@startlink[1]{}%
    \providecommand \@@endlink[0]{}%
    \providecommand \url  [0]{\begingroup\@sanitize@url \@url }%
    \providecommand \@url [1]{\endgroup\@href {#1}{\urlprefix }}%
    \providecommand \urlprefix  [0]{URL }%
    \providecommand \Eprint [0]{\href }%
    \providecommand \doibase [0]{http://dx.doi.org/}%
    \providecommand \selectlanguage [0]{\@gobble}%
    \providecommand \bibinfo  [0]{\@secondoftwo}%
    \providecommand \bibfield  [0]{\@secondoftwo}%
    \providecommand \translation [1]{[#1]}%
    \providecommand \BibitemOpen [0]{}%
    \providecommand \bibitemStop [0]{}%
    \providecommand \bibitemNoStop [0]{.\EOS\space}%
    \providecommand \EOS [0]{\spacefactor3000\relax}%
    \providecommand \BibitemShut  [1]{\csname bibitem#1\endcsname}%
    \let\auto@bib@innerbib\@empty
    \bibitem [{\citenamefont {Landau}\ and\ \citenamefont {Lifshitz}(1971)}]{LL}%
      \BibitemOpen
      \bibfield  {author} {\bibinfo {author} {\bibfnamefont {L.~D.}\ \bibnamefont
      {Landau}}\ and\ \bibinfo {author} {\bibfnamefont {E.~M.}\ \bibnamefont
      {Lifshitz}},\ }\href@noop {} {\emph {\bibinfo {title} {The Classical Theory
      of Fields}}}\ (\bibinfo  {publisher} {Pergamon Press},\ \bibinfo {address}
      {Oxford},\ \bibinfo {year} {1971})\BibitemShut {NoStop}%
    \bibitem [{\citenamefont {Ternov}(1995)}]{Ternov1995}%
      \BibitemOpen
      \bibfield  {author} {\bibinfo {author} {\bibfnamefont {I.~M.}\ \bibnamefont
      {Ternov}},\ }\href {http://stacks.iop.org/1063-7869/38/i=4/a=R03} {\bibfield
      {journal} {\bibinfo  {journal} {Physics-Uspekhi}\ }\textbf {\bibinfo {volume}
      {38}},\ \bibinfo {pages} {409} (\bibinfo {year} {1995})}\BibitemShut
      {NoStop}%
    \bibitem [{\citenamefont {Seipt}\ \emph {et~al.}(2018)\citenamefont {Seipt},
      \citenamefont {Del~Sorbo}, \citenamefont {Ridgers},\ and\ \citenamefont
      {Thomas}}]{seipt-radiative_pol-pra2018}%
      \BibitemOpen
      \bibfield  {author} {\bibinfo {author} {\bibfnamefont {D.}~\bibnamefont
      {Seipt}}, \bibinfo {author} {\bibfnamefont {D.}~\bibnamefont {Del~Sorbo}},
      \bibinfo {author} {\bibfnamefont {C.~P.}\ \bibnamefont {Ridgers}}, \ and\
      \bibinfo {author} {\bibfnamefont {A.~G.~R.}\ \bibnamefont {Thomas}},\ }\href
      {\doibase 10.1103/PhysRevA.98.023417} {\bibfield  {journal} {\bibinfo
      {journal} {Phys. Rev. A}\ }\textbf {\bibinfo {volume} {98}},\ \bibinfo
      {pages} {023417} (\bibinfo {year} {2018})}\BibitemShut {NoStop}%
    \bibitem [{\citenamefont {Li}\ \emph {et~al.}(2018)\citenamefont {Li},
      \citenamefont {Shaisultanov}, \citenamefont {Hatsagortsyan}, \citenamefont
      {Wan}, \citenamefont {Keitel},\ and\ \citenamefont
      {Li}}]{liyanfei-arxiv2018}%
      \BibitemOpen
      \bibfield  {author} {\bibinfo {author} {\bibfnamefont {Y.-F.}\ \bibnamefont
      {Li}}, \bibinfo {author} {\bibfnamefont {R.}~\bibnamefont {Shaisultanov}},
      \bibinfo {author} {\bibfnamefont {K.~Z.}\ \bibnamefont {Hatsagortsyan}},
      \bibinfo {author} {\bibfnamefont {F.}~\bibnamefont {Wan}}, \bibinfo {author}
      {\bibfnamefont {C.~H.}\ \bibnamefont {Keitel}}, \ and\ \bibinfo {author}
      {\bibfnamefont {J.-X.}\ \bibnamefont {Li}},\ }\href@noop {} {\  (\bibinfo
      {year} {2018})},\ \Eprint {http://arxiv.org/abs/1812.07229} {arXiv:1812.07229
      [physics.plasm-ph]} \BibitemShut {NoStop}%
    \bibitem [{\citenamefont {Wen}\ \emph {et~al.}(2017)\citenamefont {Wen},
      \citenamefont {Keitel},\ and\ \citenamefont {Bauke}}]{wenmeng-pra2017}%
      \BibitemOpen
      \bibfield  {author} {\bibinfo {author} {\bibfnamefont {M.}~\bibnamefont
      {Wen}}, \bibinfo {author} {\bibfnamefont {C.~H.}\ \bibnamefont {Keitel}}, \
      and\ \bibinfo {author} {\bibfnamefont {H.}~\bibnamefont {Bauke}},\ }\href
      {\doibase 10.1103/PhysRevA.95.042102} {\bibfield  {journal} {\bibinfo
      {journal} {Physical Review A}\ }\textbf {\bibinfo {volume} {95}} (\bibinfo
      {year} {2017}),\ 10.1103/PhysRevA.95.042102}\BibitemShut {NoStop}%
    \bibitem [{\citenamefont {Shen}\ \emph {et~al.}(2018)\citenamefont {Shen},
      \citenamefont {Bu}, \citenamefont {Xu}, \citenamefont {Xu}, \citenamefont
      {Ji}, \citenamefont {Li},\ and\ \citenamefont {Xu}}]{SEL}%
      \BibitemOpen
      \bibfield  {author} {\bibinfo {author} {\bibfnamefont {B.}~\bibnamefont
      {Shen}}, \bibinfo {author} {\bibfnamefont {Z.}~\bibnamefont {Bu}}, \bibinfo
      {author} {\bibfnamefont {J.}~\bibnamefont {Xu}}, \bibinfo {author}
      {\bibfnamefont {T.}~\bibnamefont {Xu}}, \bibinfo {author} {\bibfnamefont
      {L.}~\bibnamefont {Ji}}, \bibinfo {author} {\bibfnamefont {R.}~\bibnamefont
      {Li}}, \ and\ \bibinfo {author} {\bibfnamefont {Z.}~\bibnamefont {Xu}},\
      }\href {http://stacks.iop.org/0741-3335/60/i=4/a=044002} {\bibfield
      {journal} {\bibinfo  {journal} {Plasma Physics and Controlled Fusion}\
      }\textbf {\bibinfo {volume} {60}},\ \bibinfo {pages} {044002} (\bibinfo
      {year} {2018})}\BibitemShut {NoStop}%
    \bibitem [{\citenamefont {{Extreme Light Infrastructure European
      Project}}()}]{ELIweb}%
      \BibitemOpen
      \bibfield  {author} {\bibinfo {author} {\bibnamefont {{Extreme Light
      Infrastructure European Project}}},\ }\href {https://eli-laser.eu/} {\bibinfo
       {journal} {www.eli-laser.eu}\ }\BibitemShut {NoStop}%
    \bibitem [{\citenamefont {{Exawatt Center for Extreme Light
      Studies}}()}]{XCELSweb}%
      \BibitemOpen
    \bibfield  {journal} {  }\bibfield  {author} {\bibinfo {author} {\bibnamefont
      {{Exawatt Center for Extreme Light Studies}}},\ }\href
      {http://www.xcels.iapras.ru/} {\bibinfo  {journal} {www.xcels.iapras.ru}\
      }\BibitemShut {NoStop}%
    \bibitem [{\citenamefont {{Apollon multi-PW laser Users
      Facility}}()}]{Apollonweb}%
      \BibitemOpen
    \bibfield  {journal} {  }\bibfield  {author} {\bibinfo {author} {\bibnamefont
      {{Apollon multi-PW laser Users Facility}}},\ }\href {www.polytechnique.edu}
      {\bibinfo  {journal} {www.polytech nique.edu}\ }\BibitemShut {NoStop}%
    \bibitem [{\citenamefont {{The Vulcan 10-PW project}}()}]{Vulcanweb}%
      \BibitemOpen
    \bibfield  {journal} {  }\bibfield  {author} {\bibinfo {author} {\bibnamefont
      {{The Vulcan 10-PW project}}},\ }\href {www.clf.stfc.ac.uk} {\bibinfo
      {journal} {www.clf.stfc.ac.uk}\ }\BibitemShut {NoStop}%
    \bibitem [{\citenamefont {Li}\ \emph {et~al.}(2017{\natexlab{a}})\citenamefont
      {Li}, \citenamefont {Liang}, \citenamefont {Leng},\ and\ \citenamefont
      {Xu}}]{SULF}%
      \BibitemOpen
    \bibfield  {journal} {  }\bibfield  {author} {\bibinfo {author} {\bibfnamefont
      {R.}~\bibnamefont {Li}}, \bibinfo {author} {\bibfnamefont {X.}~\bibnamefont
      {Liang}}, \bibinfo {author} {\bibfnamefont {Y.}~\bibnamefont {Leng}}, \ and\
      \bibinfo {author} {\bibfnamefont {Z.}~\bibnamefont {Xu}},\ }in\ \href
      {http://aappsdpp.org/DPPProgramlatest/pdf/L-I9.pdf} {\emph {\bibinfo
      {booktitle} {1st AAPPS-DPP meeting}}}\ (\bibinfo {year} {2017})\ \bibinfo
      {note} {1st AAPPS-DPP meeting}\BibitemShut {NoStop}%
    \bibitem [{\citenamefont {Wistisen}\ \emph {et~al.}(2018)\citenamefont
      {Wistisen}, \citenamefont {Di~Piazza}, \citenamefont {Knudsen},\ and\
      \citenamefont {Uggerhoj}}]{wistisen-NC2018}%
      \BibitemOpen
      \bibfield  {author} {\bibinfo {author} {\bibfnamefont {T.~N.}\ \bibnamefont
      {Wistisen}}, \bibinfo {author} {\bibfnamefont {A.}~\bibnamefont {Di~Piazza}},
      \bibinfo {author} {\bibfnamefont {H.~V.}\ \bibnamefont {Knudsen}}, \ and\
      \bibinfo {author} {\bibfnamefont {U.~I.}\ \bibnamefont {Uggerhoj}},\ }\href
      {\doibase 10.1038/s41467-018-03165-4} {\bibfield  {journal} {\bibinfo
      {journal} {Nature Communications}\ }\textbf {\bibinfo {volume} {9}},\
      (\bibinfo {year} {2018})}\BibitemShut {NoStop}%
    \bibitem [{\citenamefont {Cole}\ \emph {et~al.}(2018)\citenamefont {Cole},
      \citenamefont {Behm}, \citenamefont {Gerstmayr}, \citenamefont {Blackburn},
      \citenamefont {Wood}, \citenamefont {Baird}, \citenamefont {Duff},
      \citenamefont {Harvey}, \citenamefont {Ilderton}, \citenamefont {Joglekar},
      \citenamefont {Krushelnick}, \citenamefont {Kuschel}, \citenamefont
      {Marklund}, \citenamefont {McKenna}, \citenamefont {Murphy}, \citenamefont
      {Poder}, \citenamefont {Ridgers}, \citenamefont {Samarin}, \citenamefont
      {Sarri}, \citenamefont {Symes}, \citenamefont {Thomas}, \citenamefont
      {Warwick}, \citenamefont {Zepf}, \citenamefont {Najmudin},\ and\
      \citenamefont {Mangles}}]{cole-PRX2018}%
      \BibitemOpen
      \bibfield  {author} {\bibinfo {author} {\bibfnamefont {J.~M.}\ \bibnamefont
      {Cole}}, \bibinfo {author} {\bibfnamefont {K.~T.}\ \bibnamefont {Behm}},
      \bibinfo {author} {\bibfnamefont {E.}~\bibnamefont {Gerstmayr}}, \bibinfo
      {author} {\bibfnamefont {T.~G.}\ \bibnamefont {Blackburn}}, \bibinfo {author}
      {\bibfnamefont {J.~C.}\ \bibnamefont {Wood}}, \bibinfo {author}
      {\bibfnamefont {C.~D.}\ \bibnamefont {Baird}}, \bibinfo {author}
      {\bibfnamefont {M.~J.}\ \bibnamefont {Duff}}, \bibinfo {author}
      {\bibfnamefont {C.}~\bibnamefont {Harvey}}, \bibinfo {author} {\bibfnamefont
      {A.}~\bibnamefont {Ilderton}}, \bibinfo {author} {\bibfnamefont {A.~S.}\
      \bibnamefont {Joglekar}}, \bibinfo {author} {\bibfnamefont {K.}~\bibnamefont
      {Krushelnick}}, \bibinfo {author} {\bibfnamefont {S.}~\bibnamefont
      {Kuschel}}, \bibinfo {author} {\bibfnamefont {M.}~\bibnamefont {Marklund}},
      \bibinfo {author} {\bibfnamefont {P.}~\bibnamefont {McKenna}}, \bibinfo
      {author} {\bibfnamefont {C.~D.}\ \bibnamefont {Murphy}}, \bibinfo {author}
      {\bibfnamefont {K.}~\bibnamefont {Poder}}, \bibinfo {author} {\bibfnamefont
      {C.~P.}\ \bibnamefont {Ridgers}}, \bibinfo {author} {\bibfnamefont {G.~M.}\
      \bibnamefont {Samarin}}, \bibinfo {author} {\bibfnamefont {G.}~\bibnamefont
      {Sarri}}, \bibinfo {author} {\bibfnamefont {D.~R.}\ \bibnamefont {Symes}},
      \bibinfo {author} {\bibfnamefont {A.~G.~R.}\ \bibnamefont {Thomas}}, \bibinfo
      {author} {\bibfnamefont {J.}~\bibnamefont {Warwick}}, \bibinfo {author}
      {\bibfnamefont {M.}~\bibnamefont {Zepf}}, \bibinfo {author} {\bibfnamefont
      {Z.}~\bibnamefont {Najmudin}}, \ and\ \bibinfo {author} {\bibfnamefont
      {S.~P.~D.}\ \bibnamefont {Mangles}},\ }\href {\doibase
      10.1103/PhysRevX.8.011020} {\bibfield  {journal} {\bibinfo  {journal}
      {Physical Review X}\ }\textbf {\bibinfo {volume} {8}},\ \bibinfo {pages} {11}
      (\bibinfo {year} {2018})}\BibitemShut {NoStop}%
    \bibitem [{\citenamefont {Poder}\ \emph {et~al.}(2018)\citenamefont {Poder},
      \citenamefont {Tamburini}, \citenamefont {Sarri}, \citenamefont {Di~Piazza},
      \citenamefont {Kuschel}, \citenamefont {Baird}, \citenamefont {Behm},
      \citenamefont {Bohlen}, \citenamefont {Cole}, \citenamefont {Corvan},
      \citenamefont {Duff}, \citenamefont {Gerstmayr}, \citenamefont {Keitel},
      \citenamefont {Krushelnick}, \citenamefont {Mangles}, \citenamefont
      {McKenna}, \citenamefont {Murphy}, \citenamefont {Najmudin}, \citenamefont
      {Ridgers}, \citenamefont {Samarin}, \citenamefont {Symes}, \citenamefont
      {Thomas}, \citenamefont {Warwick},\ and\ \citenamefont
      {Zepf}}]{poder-PRX2018}%
      \BibitemOpen
      \bibfield  {author} {\bibinfo {author} {\bibfnamefont {K.}~\bibnamefont
      {Poder}}, \bibinfo {author} {\bibfnamefont {M.}~\bibnamefont {Tamburini}},
      \bibinfo {author} {\bibfnamefont {G.}~\bibnamefont {Sarri}}, \bibinfo
      {author} {\bibfnamefont {A.}~\bibnamefont {Di~Piazza}}, \bibinfo {author}
      {\bibfnamefont {S.}~\bibnamefont {Kuschel}}, \bibinfo {author} {\bibfnamefont
      {C.~D.}\ \bibnamefont {Baird}}, \bibinfo {author} {\bibfnamefont
      {K.}~\bibnamefont {Behm}}, \bibinfo {author} {\bibfnamefont {S.}~\bibnamefont
      {Bohlen}}, \bibinfo {author} {\bibfnamefont {J.~M.}\ \bibnamefont {Cole}},
      \bibinfo {author} {\bibfnamefont {D.~J.}\ \bibnamefont {Corvan}}, \bibinfo
      {author} {\bibfnamefont {M.}~\bibnamefont {Duff}}, \bibinfo {author}
      {\bibfnamefont {E.}~\bibnamefont {Gerstmayr}}, \bibinfo {author}
      {\bibfnamefont {C.~H.}\ \bibnamefont {Keitel}}, \bibinfo {author}
      {\bibfnamefont {K.}~\bibnamefont {Krushelnick}}, \bibinfo {author}
      {\bibfnamefont {S.~P.~D.}\ \bibnamefont {Mangles}}, \bibinfo {author}
      {\bibfnamefont {P.}~\bibnamefont {McKenna}}, \bibinfo {author} {\bibfnamefont
      {C.~D.}\ \bibnamefont {Murphy}}, \bibinfo {author} {\bibfnamefont
      {Z.}~\bibnamefont {Najmudin}}, \bibinfo {author} {\bibfnamefont {C.~P.}\
      \bibnamefont {Ridgers}}, \bibinfo {author} {\bibfnamefont {G.~M.}\
      \bibnamefont {Samarin}}, \bibinfo {author} {\bibfnamefont {D.~R.}\
      \bibnamefont {Symes}}, \bibinfo {author} {\bibfnamefont {A.~G.~R.}\
      \bibnamefont {Thomas}}, \bibinfo {author} {\bibfnamefont {J.}~\bibnamefont
      {Warwick}}, \ and\ \bibinfo {author} {\bibfnamefont {M.}~\bibnamefont
      {Zepf}},\ }\href {\doibase 10.1103/PhysRevX.8.031004} {\bibfield  {journal}
      {\bibinfo  {journal} {Phys. Rev. X}\ }\textbf {\bibinfo {volume} {8}},\
      \bibinfo {pages} {031004} (\bibinfo {year} {2018})}\BibitemShut {NoStop}%
    \bibitem [{\citenamefont {Harvey}\ \emph {et~al.}(2017)\citenamefont {Harvey},
      \citenamefont {Gonoskov}, \citenamefont {Ilderton},\ and\ \citenamefont
      {Marklund}}]{Quenching}%
      \BibitemOpen
      \bibfield  {author} {\bibinfo {author} {\bibfnamefont {C.}~\bibnamefont
      {Harvey}}, \bibinfo {author} {\bibfnamefont {A.}~\bibnamefont {Gonoskov}},
      \bibinfo {author} {\bibfnamefont {A.}~\bibnamefont {Ilderton}}, \ and\
      \bibinfo {author} {\bibfnamefont {M.}~\bibnamefont {Marklund}},\ }\href
      {https://link.aps.org/doi/10.1103/PhysRevLett.118.105004} {\bibfield
      {journal} {\bibinfo  {journal} {Physical Review Letters}\ }\textbf {\bibinfo
      {volume} {118}},\ \bibinfo {pages} {105004} (\bibinfo {year}
      {2017})}\BibitemShut {NoStop}%
    \bibitem [{\citenamefont {Geng}\ \emph {et~al.}(2018)\citenamefont {Geng},
      \citenamefont {Ji}, \citenamefont {Shen}, \citenamefont {Feng}, \citenamefont
      {Guo}, \citenamefont {Yu}, \citenamefont {Zhang},\ and\ \citenamefont
      {Xu}}]{Geng-arxiv2018}%
      \BibitemOpen
      \bibfield  {author} {\bibinfo {author} {\bibfnamefont {X.~S.}\ \bibnamefont
      {Geng}}, \bibinfo {author} {\bibfnamefont {L.~L.}\ \bibnamefont {Ji}},
      \bibinfo {author} {\bibfnamefont {B.~F.}\ \bibnamefont {Shen}}, \bibinfo
      {author} {\bibfnamefont {B.}~\bibnamefont {Feng}}, \bibinfo {author}
      {\bibfnamefont {Z.}~\bibnamefont {Guo}}, \bibinfo {author} {\bibfnamefont
      {Q.}~\bibnamefont {Yu}}, \bibinfo {author} {\bibfnamefont {L.~G.}\
      \bibnamefont {Zhang}}, \ and\ \bibinfo {author} {\bibfnamefont {Z.~Z.}\
      \bibnamefont {Xu}},\ }\href@noop {} {\  (\bibinfo {year} {2018})},\ \Eprint
      {http://arxiv.org/abs/1811.04741} {arXiv:1811.04741 [physics.plasm-ph]}
      \BibitemShut {NoStop}%
    \bibitem [{\citenamefont {Gonoskov}\ \emph {et~al.}(2015)\citenamefont
      {Gonoskov}, \citenamefont {Bastrakov}, \citenamefont {Efimenko},
      \citenamefont {Ilderton}, \citenamefont {Marklund}, \citenamefont {Meyerov},
      \citenamefont {Muraviev}, \citenamefont {Sergeev}, \citenamefont {Surmin},\
      and\ \citenamefont {Wallin}}]{PICQED}%
      \BibitemOpen
      \bibfield  {author} {\bibinfo {author} {\bibfnamefont {A.}~\bibnamefont
      {Gonoskov}}, \bibinfo {author} {\bibfnamefont {S.}~\bibnamefont {Bastrakov}},
      \bibinfo {author} {\bibfnamefont {E.}~\bibnamefont {Efimenko}}, \bibinfo
      {author} {\bibfnamefont {A.}~\bibnamefont {Ilderton}}, \bibinfo {author}
      {\bibfnamefont {M.}~\bibnamefont {Marklund}}, \bibinfo {author}
      {\bibfnamefont {I.}~\bibnamefont {Meyerov}}, \bibinfo {author} {\bibfnamefont
      {A.}~\bibnamefont {Muraviev}}, \bibinfo {author} {\bibfnamefont
      {A.}~\bibnamefont {Sergeev}}, \bibinfo {author} {\bibfnamefont
      {I.}~\bibnamefont {Surmin}}, \ and\ \bibinfo {author} {\bibfnamefont
      {E.}~\bibnamefont {Wallin}},\ }\href {\doibase 10.1103/PhysRevE.92.023305}
      {\bibfield  {journal} {\bibinfo  {journal} {Phys Rev E Stat Nonlin Soft
      Matter Phys}\ }\textbf {\bibinfo {volume} {92}},\ \bibinfo {pages} {023305}
      (\bibinfo {year} {2015})}\BibitemShut {NoStop}%
    \bibitem [{\citenamefont {Del~Sorbo}\ \emph {et~al.}(2018)\citenamefont
      {Del~Sorbo}, \citenamefont {Seipt}, \citenamefont {Thomas},\ and\
      \citenamefont {Ridgers}}]{sorbo-pol-ppcf2018}%
      \BibitemOpen
      \bibfield  {author} {\bibinfo {author} {\bibfnamefont {D.}~\bibnamefont
      {Del~Sorbo}}, \bibinfo {author} {\bibfnamefont {D.}~\bibnamefont {Seipt}},
      \bibinfo {author} {\bibfnamefont {A.~G.~R.}\ \bibnamefont {Thomas}}, \ and\
      \bibinfo {author} {\bibfnamefont {C.~P.}\ \bibnamefont {Ridgers}},\ }\href
      {\doibase 10.1088/1361-6587/aab979} {\bibfield  {journal} {\bibinfo
      {journal} {Plasma Physics and Controlled Fusion}\ }\textbf {\bibinfo {volume}
      {60}} (\bibinfo {year} {2018}),\ 10.1088/1361-6587/aab979}\BibitemShut
      {NoStop}%
    \bibitem [{\citenamefont {Neitz}\ and\ \citenamefont
      {Di~Piazza}(2013)}]{dipiazza-stochastic-prl2013}%
      \BibitemOpen
      \bibfield  {author} {\bibinfo {author} {\bibfnamefont {N.}~\bibnamefont
      {Neitz}}\ and\ \bibinfo {author} {\bibfnamefont {A.}~\bibnamefont
      {Di~Piazza}},\ }\href {\doibase 10.1103/PhysRevLett.111.054802} {\bibfield
      {journal} {\bibinfo  {journal} {Phys Rev Lett}\ }\textbf {\bibinfo {volume}
      {111}},\ \bibinfo {pages} {054802} (\bibinfo {year} {2013})}\BibitemShut
      {NoStop}%
    \bibitem [{\citenamefont {Vranic}\ \emph {et~al.}(2016)\citenamefont {Vranic},
      \citenamefont {Grismayer}, \citenamefont {Fonseca},\ and\ \citenamefont
      {Silva}}]{vranic-qrr-njp2016}%
      \BibitemOpen
      \bibfield  {author} {\bibinfo {author} {\bibfnamefont {M.}~\bibnamefont
      {Vranic}}, \bibinfo {author} {\bibfnamefont {T.}~\bibnamefont {Grismayer}},
      \bibinfo {author} {\bibfnamefont {R.~A.}\ \bibnamefont {Fonseca}}, \ and\
      \bibinfo {author} {\bibfnamefont {L.~O.}\ \bibnamefont {Silva}},\ }\href
      {\doibase 10.1088/1367-2630/18/7/073035} {\bibfield  {journal} {\bibinfo
      {journal} {New Journal of Physics}\ }\textbf {\bibinfo {volume} {18}},\
      (\bibinfo {year} {2016})}\BibitemShut {NoStop}%
    \bibitem [{\citenamefont {Li}\ \emph {et~al.}(2017{\natexlab{b}})\citenamefont
      {Li}, \citenamefont {Chen}, \citenamefont {Hatsagortsyan},\ and\
      \citenamefont {Keitel}}]{lijianxing-angle_resolved_pho-scirep2017}%
      \BibitemOpen
      \bibfield  {author} {\bibinfo {author} {\bibfnamefont {J.~X.}\ \bibnamefont
      {Li}}, \bibinfo {author} {\bibfnamefont {Y.~Y.}\ \bibnamefont {Chen}},
      \bibinfo {author} {\bibfnamefont {K.~Z.}\ \bibnamefont {Hatsagortsyan}}, \
      and\ \bibinfo {author} {\bibfnamefont {C.~H.}\ \bibnamefont {Keitel}},\
      }\href {\doibase 10.1038/s41598-017-11871-0} {\bibfield  {journal} {\bibinfo
      {journal} {Sci Rep}\ }\textbf {\bibinfo {volume} {7}},\ \bibinfo {pages}
      {11556} (\bibinfo {year} {2017}{\natexlab{b}})}\BibitemShut {NoStop}%
    \bibitem [{\citenamefont {Ridgers}\ \emph {et~al.}(2017)\citenamefont
      {Ridgers}, \citenamefont {Blackburn}, \citenamefont {Del~Sorbo},
      \citenamefont {Bradley}, \citenamefont {Slade-Lowther}, \citenamefont
      {Baird}, \citenamefont {Mangles}, \citenamefont {McKenna}, \citenamefont
      {Marklund}, \citenamefont {Murphy},\ and\ \citenamefont
      {Thomas}}]{ridgers-signature_qrr-2017}%
      \BibitemOpen
      \bibfield  {author} {\bibinfo {author} {\bibfnamefont {C.~P.}\ \bibnamefont
      {Ridgers}}, \bibinfo {author} {\bibfnamefont {T.~G.}\ \bibnamefont
      {Blackburn}}, \bibinfo {author} {\bibfnamefont {D.}~\bibnamefont
      {Del~Sorbo}}, \bibinfo {author} {\bibfnamefont {L.~E.}\ \bibnamefont
      {Bradley}}, \bibinfo {author} {\bibfnamefont {C.}~\bibnamefont
      {Slade-Lowther}}, \bibinfo {author} {\bibfnamefont {C.~D.}\ \bibnamefont
      {Baird}}, \bibinfo {author} {\bibfnamefont {S.~P.~D.}\ \bibnamefont
      {Mangles}}, \bibinfo {author} {\bibfnamefont {P.}~\bibnamefont {McKenna}},
      \bibinfo {author} {\bibfnamefont {M.}~\bibnamefont {Marklund}}, \bibinfo
      {author} {\bibfnamefont {C.~D.}\ \bibnamefont {Murphy}}, \ and\ \bibinfo
      {author} {\bibfnamefont {A.~G.~R.}\ \bibnamefont {Thomas}},\ }\href {\doibase
      10.1017/s0022377817000642} {\bibfield  {journal} {\bibinfo  {journal}
      {Journal of Plasma Physics}\ }\textbf {\bibinfo {volume} {83}},\  (\bibinfo
      {year} {2017})}\BibitemShut {NoStop}%
    \bibitem [{\citenamefont {Breit}\ and\ \citenamefont
      {Wheeler}(1934)}]{BreitWheelerPP}%
      \BibitemOpen
      \bibfield  {author} {\bibinfo {author} {\bibfnamefont {G.}~\bibnamefont
      {Breit}}\ and\ \bibinfo {author} {\bibfnamefont {J.~A.}\ \bibnamefont
      {Wheeler}},\ }\href {https://link.aps.org/doi/10.1103/PhysRev.46.1087}
      {\bibfield  {journal} {\bibinfo  {journal} {Physical Review}\ }\textbf
      {\bibinfo {volume} {46}},\ \bibinfo {pages} {1087} (\bibinfo {year}
      {1934})}\BibitemShut {NoStop}%
    \bibitem [{\citenamefont {Ritus}(1985)}]{Ritus}%
      \BibitemOpen
      \bibfield  {author} {\bibinfo {author} {\bibfnamefont {V.}~\bibnamefont
      {Ritus}},\ }\href@noop {} {\bibfield  {journal} {\bibinfo  {journal} {J.
      Russ. Laser Res.}\ }\textbf {\bibinfo {volume} {6}},\ \bibinfo {pages} {497}
      (\bibinfo {year} {1985})}\BibitemShut {NoStop}%
    \bibitem [{Note1()}]{Note1}%
      \BibitemOpen
      \bibinfo {note} {$\Delta p_x$ is used for convenience.}\BibitemShut {Stop}%
    \bibitem [{\citenamefont {Salamin}\ and\ \citenamefont
      {Keitel}(2002)}]{SalaminTightlyfocused}%
      \BibitemOpen
      \bibfield  {author} {\bibinfo {author} {\bibfnamefont {Y.~I.}\ \bibnamefont
      {Salamin}}\ and\ \bibinfo {author} {\bibfnamefont {C.~H.}\ \bibnamefont
      {Keitel}},\ }\href {\doibase 10.1103/PhysRevLett.88.095005} {\bibfield
      {journal} {\bibinfo  {journal} {Phys Rev Lett}\ }\textbf {\bibinfo {volume}
      {88}},\ \bibinfo {pages} {095005} (\bibinfo {year} {2002})}\BibitemShut
      {NoStop}%
    \bibitem [{\citenamefont {Jackson}(1999)}]{Jackson}%
      \BibitemOpen
      \bibfield  {author} {\bibinfo {author} {\bibfnamefont {J.~D.}\ \bibnamefont
      {Jackson}},\ }\href@noop {} {\emph {\bibinfo {title} {Classical
      Electrodynamics (3rd ed.)}}}\ (\bibinfo  {publisher} {John Wiley \& Sons},\
      \bibinfo {address} {New York},\ \bibinfo {year} {1999})\BibitemShut {NoStop}%
    \bibitem [{\citenamefont {Schwinger}(1948)}]{schwinger-anomalous}%
      \BibitemOpen
      \bibfield  {author} {\bibinfo {author} {\bibfnamefont {J.}~\bibnamefont
      {Schwinger}},\ }\href {\doibase 10.1103/PhysRev.73.416} {\bibfield  {journal}
      {\bibinfo  {journal} {Phys. Rev.}\ }\textbf {\bibinfo {volume} {73}},\
      \bibinfo {pages} {416} (\bibinfo {year} {1948})}\BibitemShut {NoStop}%
    \end{thebibliography}
\end{document}